\documentclass{iau} 
\usepackage{graphicx}
\usepackage{amsmath}
\usepackage{natbib}
\usepackage{aas_macros}

\newcommand{\gpcm}{\, \mathrm{g} \, \mathrm{cm}^{-3}}

\defcitealias{Skinner_2013}{SO13}
\defcitealias{Jiang_2012}{J12}

\title[\texttt{VETTAM}] 
{A radiation hydrodynamics scheme on adaptive meshes using the Variable Eddington Tensor (VET) closure}

\author[S.H.~Menon et al.]   
{Shyam H.~Menon$^1$, Christoph~Federrath$^{1,2}$
 , Mark R. Krumholz$^{1,2}$, \\ Rolf~Kuiper$^3$, Benjamin D.~Wibking$^{1,2}$ \and Manuel Jung$^4$}

\affiliation{
$^1$Research School of Astronomy and Astrophysics, Australian National University, Canberra, ACT~2611, Australia\\ 
$^{2}$ARC Centre of Excellence for Astronomy in Three Dimensions (ASTRO-3D), Canberra, ACT~2611, Australia\\
$^{3}$Zentrum f{\"u}r Astronomie der Universit{\"a}t Heidelberg, Institut f{\"u}r Theoretische Astrophysik, Albert-Ueberle-Stra{\ss}e~2, 69120~Heidelberg, Germany\\
$^{4}$Hamburger Sternwarte, Universit{\"a}t Hamburg, Gojenbergsweg~112, 21029~Hamburg, Germany

email: {\tt shyam.menon@anu.edu.au} \\[\affilskip]
}

\pubyear{2022}
\volume{362}  
\jname{The predictive power of computational astrophysics as a discovery tool}
\editors{D. Bisikalo, T. Hanawa, C. Boily \& J. Stone, eds.}
\begin{document}

\maketitle

\begin{abstract}
We present a new algorithm to solve the equations of radiation hydrodynamics (RHD) in a frequency-integrated, two-moment formulation. Novel features of the algorithm include i) the adoption of a non-local Variable Eddington Tensor (VET) closure for the radiation moment equations, computed with a ray-tracing method, ii) support for adaptive mesh refinement (AMR), iii) use of a time-implicit Godunov method for the hyperbolic transport of radiation, and iv) a fixed-point Picard iteration scheme to accurately handle the stiff nonlinear gas-radiation energy exchange. Tests demonstrate that our scheme works correctly, yields accurate rates of energy and momentum transfer between gas and radiation, and obtains the correct radiation field distribution even in situations where more commonly used -- but less accurate -- closure relations like the Flux-limited Diffusion and Moment-1 approximations fail. Our scheme presents an important step towards performing RHD simulations with increasing spatial and directional accuracy, effectively improving their predictive capabilities. 
\end{abstract}

\firstsection 
\section{Introduction}

Radiation hydrodynamics (RHD) plays a crucial role in the evolution of several astrophysical systems from stellar atmospheres \citep[e.g.,][]{Mihalas_1978} to cosmological reionization \citep{Gnedin_1997}. There has been significant progress in recent years to develop the numerical algorithms that are required to solve the stiff, coupled equations that govern these systems \citep[see,][for recent reviews]{Dale_2015,Teyssier_2019}. There are well-known fundamental difficulties associated with numerically solving the RHD equations, one of which is the multidimensional nature of the radiation intensity -- a function of space, time, frequency, and angular direction -- that is governed by the radiative transfer equation. A common approach to circumvent this difficulty is to integrate the RT equation over all frequencies and angles to obtain the gray radiation moment equations, reducing the dimensionality of the RHD system \citep{Mihalas_1982,Castor_2004}. However, this introduces the need for an extra closure equation to estimate the moments of the radiation intensity whose evolution is not explicitly computed. 

One commonly-used approximate closure is the flux-limited diffusion (FLD) method, which closes the equations at the first moment (the radiation flux), and uses the Eddington approximation, i.e., the assumption that the Eddington tensor is locally isotropic \citep{Levermore_1981}. However, this method often suffers from inaccuracies in the optically thin regime, or when a mixture of low- and high-opacity gas is present. A more accurate closure scheme that has recently been adopted widely is the $M_1$ closure \citep[e.g.,][]{Skinner_2013,Wibking_2021}, which adopts a local closure relation for the radiation pressure tensor, or equivalently the Eddington tensor, in terms of the local radiation energy density and flux \citep{Levermore_1984}. While the $M_1$ closure can handle transitions in optical depths for a single beam of radiation, it fails for other non-trivial geometrical distributions of radiation sources. 

In the algorithm described in this paper, we use the so-called Variable Eddington Tensor (VET) scheme \citep[e.g.,][]{Stone_1992,Jiang_2012}, a non-local scheme that does not adopt a closure relation or model \textit{a priori}, but rather computes the Eddington tensor self-consistently through a formal solution of the time-independent RT equation along discrete rays using a ray-tracing approach \citep[e.g.,][]{Davis_2012}. The self-consistently computed closure is combined with the radiation moment equations to solve for the radiation quantities. While more computationally expensive due to the required non-local ray-trace solution and its associated communication overheads, the VET approach does not face the shortcomings of the more approximate closure models discussed above \citep{Jiang_2012}. Below, we describe the equations solved by our scheme, which we couple to the \texttt{FLASH} (magneto-)hydrodynamics code, outline the brief features of our algorithm, and conclude by demonstrating the novel advantages offered by our scheme. The full details of the algorithm are provided in our recent publication \citep{Menon_2022}.

\section{Equations}
We solve the equations of non-relativistic gray (frequency-integrated) RHD in conservative form, written in the mixed-frame formulation, i.e., where the moments of the radiation intensity are written in the lab frame, and the opacities are written in the comoving frame \citep[e.g.,][]{Mihalas_1982}. We neglect scattering for simplicity, and assume the matter is always in local thermodynamic equilibrium (LTE), and treat the material property coefficients as isotropic in the comoving frame. The equations solved in our scheme are the equations of (magneto-)hydrodynamics along with the radiation moment equations, given by 
\begin{gather}
    \label{eq:radenergy}
	\frac{\partial E_r}{\partial t} + \nabla \cdot \mathbf{F}_r = -cG^0\\
	\frac{\partial \mathbf{F}_r}{\partial t} + \nabla \cdot (c^2E_r\mathbb{T}) = -c^2\mathbf{G},
    \label{eq:radflux}
\end{gather}
and
\begin{equation}
    \label{eq:G0}
    \begin{aligned}
    G^0 =& \rho \kappa_E E_r - \rho \kappa_P a_rT^4 + \rho \left(\kappa_F - 2\kappa_E \right) \frac{\mathbf{v} \cdot \mathbf{F}_r}{c^2} \\
    &+ \rho \left( \kappa_E - \kappa_F \right) \left[\frac{v^2}{c^2}E_r + \frac{\mathbf{v}\mathbf{v}}{c^2} :\mathbb{P}_r  \right] ,
\end{aligned}
\end{equation}
and
\begin{equation}
\label{eq:G}
\mathbf{G} = \rho \kappa_R \frac{\mathbf{F}_r}{c} - \rho \kappa_R E_r\frac{\mathbf{v}}{c} \cdot (\mathbb{I} + \mathbb{T}),
\end{equation}
are the time-like and space-like parts of the specific radiation four-force density for a direction-independent flux spectrum to leading order in all regimes. Source terms of magnitude $cG^0$ and $\mathbf{G}$ are added to the gas energy and energy momentum equations, respectively, to ensure the conservation of total (radiation + gas) energy and momentum. 

In the above equations $\rho$ is the mass density, $\mathbf{v}$ the gas velocity, $T$ the gas temperature, $\mathbb{I}$ the identity matrix, $c$ the speed of light in vacuum, and $a_r$ the radiation constant. $E_r$ is the lab-frame radiation energy density, $\mathbf{F}_r$ the lab-frame radiation momentum density, and $\mathbb{P}_r$ is the lab-frame radiation pressure tensor. These represent the zeroth, first and second (gray) angular moments of the radiation intensity, respectively. The material opacities $\kappa_E$ and $\kappa_F$ are the gray energy- and flux-mean opacities in the comoving frame, which we set equal to $\kappa_P$, the Planck mean opacity, and $\kappa_R$, the Rosseland mean opacity, respectively. The radiation closure relation is used to close the above system of equations, and is of the form
\begin{equation}
\label{eq:closure}
\mathbb{P}_r = \mathbb{T}E_r,
\end{equation}
where $\mathbb{T}$ is the Eddington Tensor, given by the relation between the second and zeroth moments of the gray radiation intensity $I_r(\hat{\mathbf{n}}_k)$ travelling in direction $\hat{\mathbf{n}}_k$. $I_r$ is computed independently from a formal solution of the time-independent radiative transfer equation
\begin{equation}
    \label{eq:RTeq}
    \frac{\partial I_r}{\partial s} = \kappa_P(S - I_r),
\end{equation}
where $S$ is the source function, which, for the purposes of modelling the emission from dust grains, we set equal to the frequency-integrated Planck function $B(T) = ca_RT^4/(4\pi)$. This independent solution is used to compute the angular moments $\mathbb{P}_r$ and $E_r$, to obtain the corresponding $ \mathbb{T}$, which is then used in the radiation moment equations. 

\section{Numerical Scheme}
In our scheme, we operator-split the hyperbolic hydrodynamic subsystem of equations from the radiation moment equations and treat the two separately. To evolve the hyperbolic transport for the gas quantities, we use the existing hydrodynamic solver capabilities in the \texttt{FLASH} code. The hyperbolic transport of $E_r$ and $\mathbf{F}_r$ is done using a piecewise constant (first-order) finite-volume Godunov method using a Harten-Lax-van Leer (HLL)-type Riemann solver, similar to the scheme described in \citet{Jiang_2012}. The hyperbolic fluxes and source terms are discretised in an implicit backward-Euler fashion, to permit the radiation quantities, which are governed by the light-crossing signal speed, to be evolved at the significantly larger hydrodynamic timescale. We also introduce a new, modified characteristic wavespeed criterion for the Riemann solver on AMR grids. Our criterion extends the condition described in \citet{Jiang_2013}, which performs an optical depth-dependent correction to obtain accurate solutions in the optically thick regime, to AMR grids. The Eddington tensor $\mathbb{T}$, required to solve the radiation moment equations, is pre-computed at the start of the radiation step using a solution of Equation~\ref{eq:RTeq}, which we obtain using the hybrid characteristics ray-tracer described in \citet{Buntemeyer_2016}. In addition, our scheme adopts a fixed-point Picard iteration method to couple the stiff radiation-gas nonlinear term. This is achieved by using an estimate for the gas temperature ($T_*$) in the radiation moment equations, based on the corresponding solution of $E_r$ and $\mathbf{F}_r$ to obtain an updated value of $T_*$ from the gas internal energy equation, and iterating the two stages to convergence. The converged values of $E_r$ and $\mathbf{F}_r$ are used to apply the radiation source terms to the gas energy and momentum densities, respectively. This series of steps summarizes the working of the algorithm in each simulation timestep. 

\section{Tests}
We verified that our scheme works correctly by performing a suite of numerical tests that have analytical/semi-analytical relations to compare with. Due to limited space, we only reproduce two tests here, which demonstrate the novel advantages offered by our scheme - i.e., i) the spatial accuracy -- with reduced computational costs -- offered by the support for AMR grids, and ii) the accuracy in the radiation field distribution for general problems, offered by the VET closure. All other tests, including those that specifically test the hyperbolic transport in different optical depth regimes, radiation-gas energy exchange, and other standard tests, are outlined in Menon et al.~(2022, submitted).

\subsection{Radiation Pressure Driven Shell Expansion}

\begin{figure}
    \centering
    \includegraphics[width= \textwidth]{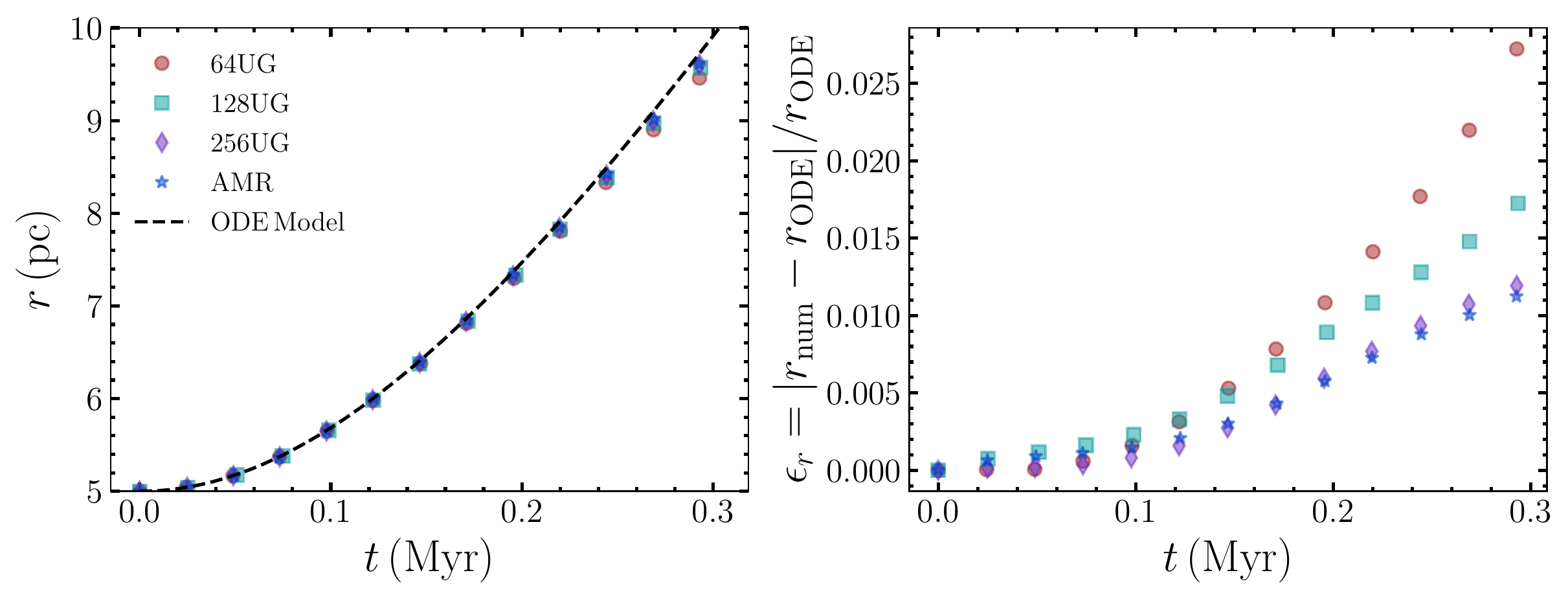}
    \caption{Numerical solution (left) for the radiation-driven thin spherical shell expansion test obtained on uniform grids (UG) with resolutions of $64^3$, $128^3$, and $256^3$ cells, and on an AMR grid with effective resolution of $256^3$ cells. Dashed lines indicate the semi-analytical ODE solution for the problem given in \citet{Skinner_2013}, and the right panel shows the relative errors for the numerical runs. The simulations are converging with increasing resolution. All errors are $<3\%$ for all relevant times, and the AMR solution is comparable to the $256^3$ UG solution, but required 30\% less computational time.}
    \label{fig:thinshellsoln}
\end{figure}

This test simulates the radiation pressure-driven expansion of a thin, dusty, spherical shell as given in \citet{Skinner_2013}. At $t=0$, a radial ($r$) density profile, representing a thin dusty shell of gas, is initialized as
\begin{equation}
    \rho_{\mathrm{sh}}(r)=\frac{M_{\mathrm{sh}}}{4 \pi r^{2} \sqrt{2 \pi R_{\mathrm{sh}}^{2}}} \exp \left(-\frac{r^{2}}{2 R_{\mathrm{sh}}^{2}}\right), 
\end{equation}
where $M_{\mathrm{sh}}$ is the gas mass in the thin shell, and $R_{\mathrm{sh}} \equiv H/(2\sqrt{2 \ln 2})$ is the half-width of the shell, where we adopt $H = 1.5 \, \mathrm{pc}$. A central radiating source is introduced, given by
\begin{equation}
    \label{eq:jstar}
    j_{*}(r)=\frac{L_{*}}{\left(2 \pi R_{*}^{2}\right)^{3 / 2}} \exp \left(-\frac{r^{2}}{2 R_{*}^{2}}\right), 
\end{equation}
where $L_*$ is the luminosity of the source and $R_*$ the size of the source. We set a value of $L_* = 1.989 \times 10^{42} \, \mathrm{erg} \, \mathrm{s}^{-1}$, $R_* = 0.625 \, \mathrm{pc}$, and a constant dust opacity of $\kappa_0 = 20 \, \mathrm{cm}^{2} \, \mathrm{g}^{-1}$. We use an isothermal equation of state for the thermal pressure, with the sound speed set to $a_0 = 2 \, \mathrm{km} \, \mathrm{s}^{-1}$, which corresponds to a gas temperature $T\sim 481 \, \mathrm{K}$, assuming a mean particle mass $\mu = m_\mathrm{H}$. The simulation is performed on a cubic domain with $x=y=z=[-10,10]\,\mathrm{pc}$, with the source at $x=0$, with outflow boundary conditions on the gas and radiation. We use AMR for this test, with a base resolution of $32^3$ grid cells, and allow up to four levels of refinement, corresponding to an effective resolution of $256^3$ grid cells. We also perform simulations on uniform grids of resolution $64^3$, $128^3$, and $256^3$ grid cells, to study the dependence of shell evolution on resolution. We compare the obtained numerical results at these resolutions in Figure~\ref{fig:thinshellsoln} with the semi-analytical ODE solution given in \citet{Skinner_2013}. We see that the numerical solution agrees with the ODE solution quite well, and increasingly so at higher resolutions. We find similar accuracy in the $256^3$ effective resolution AMR version and the $256^3$ uniform run, however, the AMR run uses about 30\% less CPU time than the uniform-grid run.

\subsection{Comparison of closure schemes: Advantage of VET}
\label{sec:shadow}

\begin{figure}
    \centering
    \includegraphics[width=\textwidth]{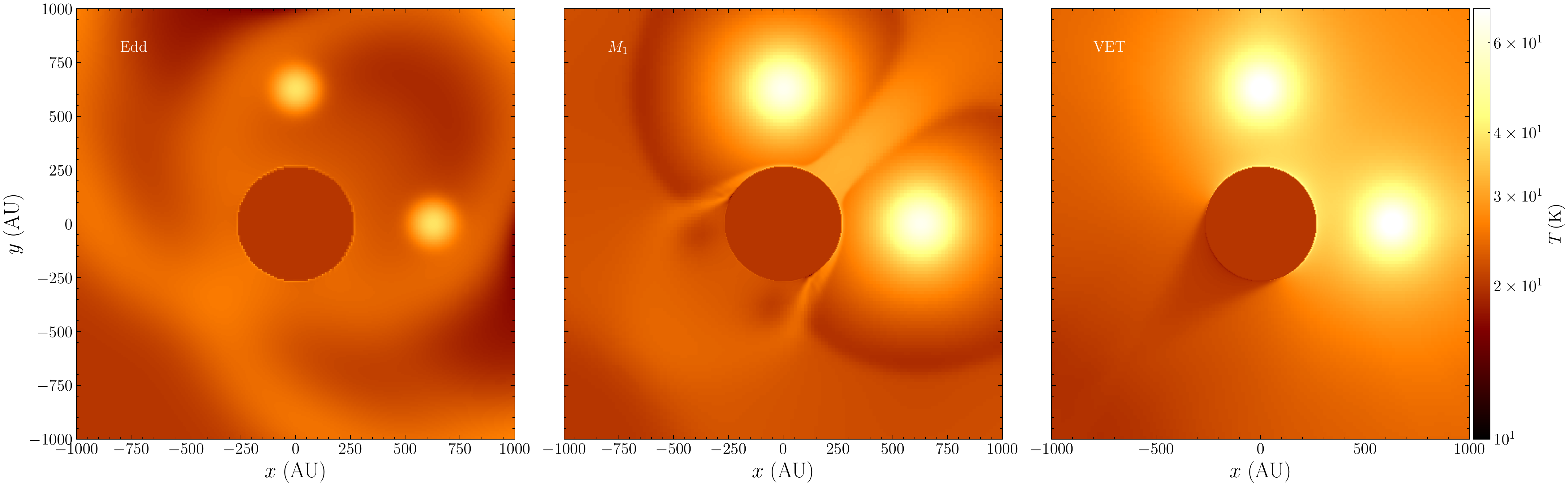}
    \caption{Comparison of the gas temperature fields obtained in the shadow test (Section~\ref{sec:shadow}) with Eddington (FLD), $M_1$ and VET closures. The simpler closure schemes produce unphysical solutions for this problem, whereas the VET reproduces the qualitatively correct solution, demonstrating the advantage of VET over other closures in semi-transparent RHD problems.}
    \label{fig:comparemethod}
\end{figure}

Here we compare our new VET approach with the more commonly adopted local closures, Eddington (used by the FLD method) and $M_1$. We set up a test where we introduce two point-like sources of radiation, modeled with the Gaussian source function $j_*(r)$ given in Equation~\ref{eq:jstar}, where we use $L_* =  10 \, L_{\odot}$ and $R_* = 54 \, \mathrm{AU}$ for both the sources. We place these sources at $90^{\deg}$ with respect to each other, at $(635,0,0) \, \mathrm{AU}$ and $(0,635,0) \, \mathrm{AU}$ respectively in a $(2000 \, \mathrm{AU})^3$ computational domain. We place a dense clump of material at the center of the domain, with radius $267 \, \mathrm{AU}$ and density $\rho_\mathrm{c} = 3.89 \times 10^{-17} \gpcm$, and an optically thin ambient medium with density $\rho_\mathrm{a} = 3.89 \times 10^{-20} \gpcm$ elsewhere. The gas temperature is spatially uniform with a value of $20 \, \mathrm{K}$, and the opacity for the radiation field is set to $\kappa_P = \kappa_R = 100 \, \mathrm{cm}^{2} \, \mathrm{g}^{-1}$. We perform three versions of this test, with only a difference of the adopted closure relation between the versions; corresponding to the Eddington (FLD) closure, the $M_1$ closure, and the VET closure, respectively. In Figure~\ref{fig:comparemethod}, we see that the Eddington and $M_1$ closures do not cast qualitatively correct shadows, whereas the VET does. This demonstrates that the VET ensures the consistent propagation of radiation in non-trivial geometrical distributions of diffuse radiation sources.

\section{Summary}
We have described an algorithm for solving the radiation hydrodynamics (RHD) equations, closed with a non-local Variable Eddington Tensor (VET), and coupled to the \texttt{FLASH} AMR (magneto-)hydrodynamics code \citep{Menon_2022}. Using numerical tests, we show that our scheme works in concert with AMR, allowing for very high resolution applications, with adequate accuracy of the radiation field. Unlike the FLD and $M_1$ closures, our VET method casts shadows as expected for complex geometrical configurations of the gas and radiation field. To our knowledge, our method is the \textit{first} VET closure-based RHD method with AMR support in the literature, and thus represents a step toward, improving the predictive capabilities of numerical RHD simulations. 

\bibliographystyle{iaulike}
\bibliography{symposium_paper}

\begin{discussion}

\discuss{Chia-Yu Hu}{The VET is expected to be more computationally expensive than the FLD and $M_1$ methods, and hence it is possible to probe higher resolution simulations with the latter class of methods. Did that play a role in why earlier methods choose these approaches?}

\discuss{Shyam}{Yes, completely agree; the VET, although more accurate, is certainly more computationally expensive than the other methods, and hence it is crucial to apply it on problems where the added accuracy could possibly matter. These represent alternate approaches to simulations -- a less accurate method, cheaper method that allows for higher resolution and further parameter space exploration vs a method with higher accuracy, that can only be used in a smaller parameter range and resolution due to its computational costs.}

\discuss{Miikka V{\"a}is{\"a}l{\"a}}{I was wondering whether you could comment on the computational performance of your scheme, and the number of angular rays required for reasonable results?}

\discuss{Shyam}{Thank you for the very good question! Although the VET method is known to be computationally expensive, its not straightforward to provide general performance statistics, since the performance is very much problem-dependent due to the use of implicit matrix inversion algorithms, whose rate convergence depends on the stiffness of the matrix. That being said, the ray-tracer is typically the bottleneck in terms of performance in our scheme, with higher angular ray resolution aggravating the performance. We find in our shadow test, however, that an angular resolution of 48~rays (using HEALPix sampling) is a reasonably good balance between performance and accuracy, and is the fiducial value we use in our scheme.}

\end{discussion}

\end{document}